\documentclass[prd,10pt,twocolumn,superscriptaddress,floatfix,nofootinbib,preprintnumbers]{revtex4-1}
\usepackage[colorlinks=true,pdfstartview=FitV,breaklinks=true]{hyperref}

\usepackage{bm}
\usepackage[normalem]{ulem}
\usepackage{graphicx}
\usepackage{multirow}
\usepackage{soul}
\usepackage{amsmath}
\usepackage{fontawesome}
\usepackage{mathrsfs}
\usepackage{amssymb}
\usepackage{yfonts}
\usepackage{cancel}
\usepackage{color}
\usepackage{xspace}
\usepackage{slashed}
\usepackage{url}
\usepackage{verbatim}
\usepackage{mathtools}
\usepackage{upgreek}
\usepackage{units}
\usepackage{siunitx}
\usepackage{amstext}
\usepackage[sc]{mathpazo}
\usepackage{booktabs}
\usepackage{tabulary}
\usepackage{tabularx}
\usepackage{etoolbox}
\usepackage[version=4]{mhchem}
\usepackage[utf8]{inputenc}

\newcommand{\keV}{\rm keV}

\newcommand{\erg}{\rm erg}
\newcommand{\s}{\rm s}

\newcommand{\MeV}{\rm MeV}

\newcommand{\bk}{\mathbf{k}}
\newcommand{\bp}{\mathbf{p}}
\newcommand{\bq}{\mathbf{q}}

\begin{document}

\title{Supernova bound on Axion-Like Particles coupled with electrons}

\author{Giuseppe Lucente}\email{giuseppe.lucente@ba.infn.it}
\affiliation{Dipartimento Interateneo di Fisica “Michelangelo Merlin”, Via Amendola 173, 70126 Bari, Italy}
\affiliation{Istituto Nazionale di Fisica Nucleare - Sezione di Bari, Via Orabona 4, 70126 Bari, Italy}%

\author{Pierluca Carenza}\email{pierluca.carenza@ba.infn.it}
\affiliation{Dipartimento Interateneo di Fisica “Michelangelo Merlin”, Via Amendola 173, 70126 Bari, Italy}
\affiliation{Istituto Nazionale di Fisica Nucleare - Sezione di Bari, Via Orabona 4, 70126 Bari, Italy}%

\date{\today}
\smallskip
\begin{abstract}
Axion-Like Particles (ALPs) coupled with electrons would be produced in a Supernova (SN) via electron-proton bremsstrahlung and electron-positron fusion. We evaluate the ALP emissivity from these processes by taking into account the ALP mass and thermal effects on electrons in the strongly degenerate and relativistic SN plasma. Using a state-of-the-art SN simulation, we evaluate the SN 1987A cooling bound on ALPs for masses in the range $1-200$~MeV, which excludes currently unprobed regions down to $g_{ae}\sim 2.5\times 10^{-10}$ at $m_a\sim 120$~MeV.
\end{abstract}
\maketitle

\section{Introduction}

\begin{figure}[t!]
\centering
\includegraphics[width=0.3\textwidth]{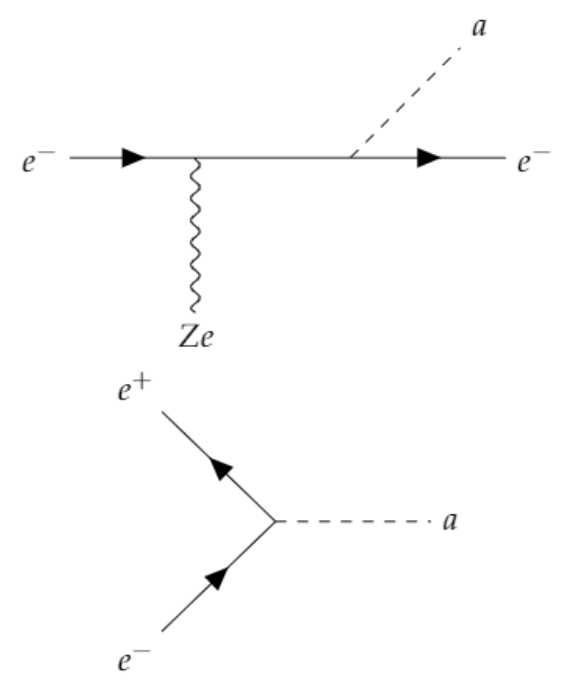}
\caption{Feynman diagrams of the electron-ion bremsstrahlung and electron-positron fusion. Note that in the electron-ion bremsstrahlung a second amplitude with the vertices interchanged is not shown.}
\label{fig:feynm}
\end{figure}

The QCD axion is a hypothetical pseudoscalar particle predicted by the Peccei-Quinn solution of the strong CP problem of the Standard Model (SM)~\cite{Peccei:1977ur,Peccei:1977hh,Weinberg:1977ma,Wilczek:1977pj}. Hadronic axions, as the minimal Kim-Shifman-Vainshtein-Zakharov (KSVZ) model, interact with hadrons and photons, while the interaction with leptons arises at further loop level~\cite{Shifman:1979if}. On the other hand, non-hadronic models, such as the Dine-Fischler-Srednicki-Zhitnitsky (DSFZ)~\cite{Dine:1981rt}, predict a tree-level axion-electron coupling, typical also of many Axion-Like Particle (ALP) models, which emerge in more general theories, as grand unified theories and string theory~\cite{Cicoli:2012sz,Kachru:2003aw,Conlon:2006ur,Choi:2006qj,Arvanitaki:2009fg}. In the following, we will consider ALPs predominantly coupled with electrons. This coupling could be probed through astrophysical arguments~\cite{Isern:2008fs,Isern:2008nt,Bertolami:2014wua,Capozzi:2020cbu,Straniero:2020iyi} and laboratory experiments~\cite{Riordan:1987aw,Bross:1989mp,Bjorken:1988as,Konaka:1986cb,Alves:2017avw,Bassompierre:1995kz,Scherdin:1991xy,Tsai:1989vw,Blumlein:1990ay,Blumlein:1991xh,Armengaud:2018cuy,LUX:2017glr,PandaX:2017ock,XENON:2020rca} (see Refs.~\cite{DiLuzio:2020wdo,Agrawal:2021dbo} for a recent review). In astrophysical context, stars in which electrons are more degenerate, such as the core of red giants (RGs) and white dwarfs (WDs) provide the most stringent bounds on the axion coupling with electrons.
Indeed, the RG bound excludes $g_{ae}\gtrsim 1.6\times10^{-13}$  \cite{Capozzi:2020cbu,Straniero:2020iyi} and the WD bound constrains $g_{ae}\gtrsim 2.8\times10^{-13}$ \cite{Isern:2008fs,Isern:2008nt,Bertolami:2014wua}. In these environments, due to the electron degeneracy, the leading ALP production channel is the electron-ion bremsstrahlung $e^{-}+Ze\rightarrow e^{-}+Ze+a$ (see upper panel of Fig.~\ref{fig:feynm}). In Ref.~\cite{Carenza:2021osu}, the ALP emission rate for this process in non-relativistic electron conditions was revised. However, a similar work is still lacking for relativistic conditions encountered in core-collapse supernovae (SNe), since more attention is required in this case. \\
A core-collapse SN is an efficient cosmic laboratory to probe ALPs and the detection of a neutrino burst from SN 1987A is a milestone in this context. Indeed, if ALPs had contributed to an excessive energy-loss in the SN core, they would have shortened the duration of the SN 1987A neutrino signal. In this way, constraints on ALPs coupled to nucleons~\cite{Raffelt:1987yt,Carenza:2019pxu,Carenza:2020cis}, photons~\cite{Lee:2018lcj,Masso:1995tw,Dolan:2017osp,Lucente:2020whw} and muons~\cite{Bollig:2020xdr,Caputo:2021rux} have been obtained (see Refs.~\cite{Calore:2020tjw,Caputo:2021kcv,Calore:2021klc} for other phenomenological consequences of the ALP production in SNe). On the other hand, in the literature, the SN 1987A cooling bound for ALPs coupled to electrons is often overlooked. However, considering a generic ALP, the coupling with electrons might be the only significant coupling with a SM particle~\cite{Dias:2014osa,Shin:1987xc,Ballesteros:2016euj} and in this case, the SN 1987A bound plays a major role~\cite{Calibbi:2020jvd}. This constraint is even more relevant for ALP masses above 1 MeV, where ALPs can decay into electron-positron pairs. Indeed, this region is probed only by beam-dump experiments~\cite{Riordan:1987aw,Bross:1989mp,Bjorken:1988as,Konaka:1986cb,Alves:2017avw,Bassompierre:1995kz,Scherdin:1991xy,Tsai:1989vw}, excluding couplings around $10^{-8}\lesssim g_{ae}\lesssim 10^{-3}$.  
The simplified approach in Ref.~\cite{Calibbi:2020jvd}\footnote{Note that the approximations used in Ref.~\cite{Calibbi:2020jvd} overestimate the ALP emissivity, as discussed in Appendix B of Ref.~\cite{Carenza:2021osu}.} gives a bound stronger than those obtained through beam-dump experiments. However, the study of ALP production in a SN via electron interactions requires a dedicated investigation because thermal plasma effects might significantly alter the ALP production rate. Indeed, the Dirac equation of electrons (and positrons) differs from the free-fermion case, affecting their dispersion relations. In the hot plasma of a SN, the fermion energy-momentum relation is modified and the electron (positron) acquires an effective mass $m_e^*\sim O(10)$~MeV~\cite{Braaten:1991hg}. In addition, in a plasma, a new quasi-particle appears, the plasmino~\cite{Braaten:1991hg}, which might partecipate in the ALP production.\\
The aim of this work is to perform the first accurate calculation of the SN emissivity of ALPs coupled to electrons. In the low ALP mass limit $m_a\lesssim O(10)$~MeV, ALPs are produced mainly via electron bremsstrahlung on the free protons in the SN core. Unexpectedly, for ALP masses larger than the effective electron mass, but lower than the SN temperature ($T\sim 30-40$~MeV), the bremsstrahlung emissivity starts to be suppressed and for $m_a\gtrsim 30$~MeV another process is found to be dominant, the electron-positron fusion $e^{+}e^{-}\rightarrow a$ (see the lower panel of Fig.~\ref{fig:feynm}), neglected in astrophysical context so far.\\
The plan of this work is as follows. In Sec.~\ref{sec:disprel} we show how the fermion dispersion relation is modified in the relativistic and degenerate SN plasma. In Sec.~\ref{sec:production} we describe the ALP production processes in the SN core. In Sec.~\ref{sec:SNbound} we calculate the SN 1987A bound on massive ALPs. Finally in Sec.~\ref{sec:conclusions} we conclude. 

\begin{figure}[t!]
\centering
\includegraphics[width=0.45\textwidth]{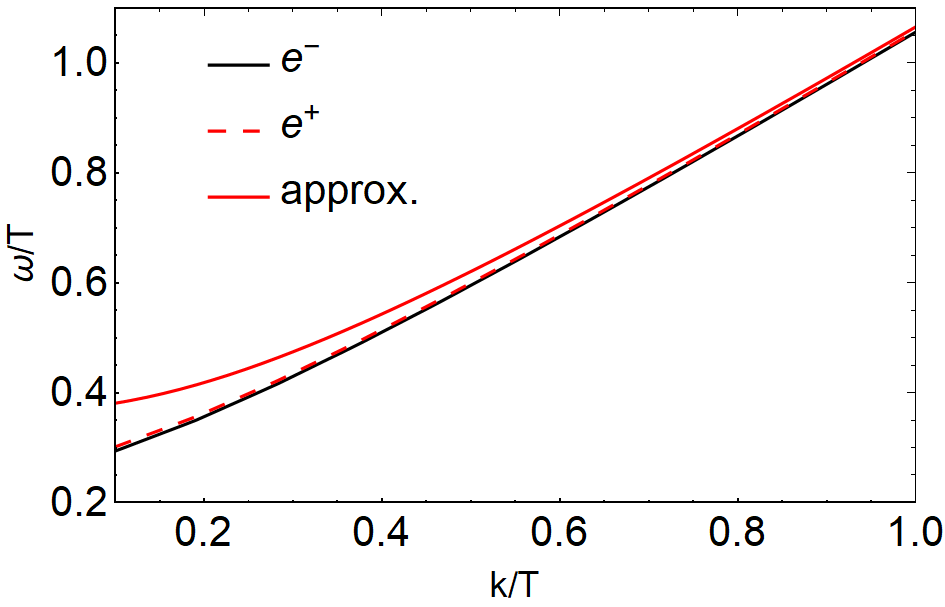}
\caption{Electron and positron dispersion relation in a SN for electron chemical potential $\mu_e=200$~MeV and temperature $T=30$~MeV. The approximated dispersion relation is a good approximation in the relativistic limit in which we are interested.}
\label{fig:disprel}
\end{figure}

\section{Impact of thermal effects on fermions and Axion-Like Particles}
\label{sec:disprel}

\subsection{Fermion dispersion relation at finite temperature}

In a SN core, the electron (positron) propagation is non trivial, since the extremely large temperature $T\sim 30-40$~MeV and the density $\rho\sim O(10^{14})$~g cm$^{-3}$ make the plasma ultra-relativistic and degenerate. At finite temperature, a plasma characterized by the fluid velocity four-vector $u^{\mu}$, modifies the structure the Dirac equation~\cite{Petitgirard:1991mf,Braaten:1991hg,Morales:1999ia}, without violating the Lorentz invariance~\cite{Weldon:1982aq}. In the plasma rest frame $u^{\mu}=(1,0,0,0)$~\cite{Weldon:1982aq}.

In a ultra-relativistic and degenerate plasma, the Dirac equation for an electron with a bare mass $m_{e}$ is \cite{Petitgirard:1991mf}\footnote{Eq.~\eqref{eq:dirac} is the only Lorentz-invariant generalization of the Dirac equation including the fluid velocity four-vector $u^{\mu}$ and the particle four-momentum operator $i\,\partial_\mu$. This reduces to the free Dirac equation in the vacuum \cite{Weldon:1989ys,Petitgirard:1991mf}.}
\begin{equation}
    ((1+A)i\slashed{\partial}+B\slashed{u}-m_{e}(1-C))\psi = 0\,,
\label{eq:dirac}
\end{equation}
where $A,\,B,\,C$ are complex numbers related to the thermal loop inside the electron propagator. The spinors associated with this equation are explicitly calculated in \cite{Weldon:1989ys}. 

The modified dispersion relation is determined by the pole of the following propagator
\begin{equation}
    S=i\frac{(1+A)\slashed{K}+B\slashed{u}+m_{e}(1-C)}{(1+A)^2K^2+2(1+A)B\,(K\cdot u)+B^2-m_{e}^2(1-C)^2}\,,
\end{equation}
where $K^\mu=(\omega,\textbf{k})$ and $u^{\mu}=(1,0,0,0)$ in the fluid rest-frame velocity. 

Plasma effects on the electron propagation give rise to a dispersion relation with a non-trivial momentum dependence which must be taken into account in any calculation in the plasma. It is also necessary to consider the residue of the pole, which accounts for the strength with which the particle is excited by the electron field and therefore the strength with which it couples to other fields. The transcendental equation 
\begin{equation}
\begin{split}
    (1+A)^2(\omega^{2}-|\bk|^{2})&+2(1+A)B\,\omega+\\
    &+B^2-m_{e}^2(1-C)^2=0\;,
\end{split}
\label{eq:trasceq}    
\end{equation}
 must be solved to determine the particle dispersion relations. This equation is solved in full generality with the $A, B, C$ functions determined by
following Ref.~\cite{Petitgirard:1991mf,Morales:1999ia}, where an explicit expression for $A,\,B,\,C$ can be found:
\begin{equation}
\begin{split}
    &A=\frac{1}{k^2}\left(\frac{1}{4}Tr[\slashed{K}Re(\Sigma')]-\omega\frac{1}{4}Tr[\slashed{u}Re(\Sigma')]\right)\,,\\
    &B=\left(\frac{\omega^2}{k^2}-1\right)\omega\frac{1}{4}Tr[\slashed{u}Re(\Sigma')]-\frac{\omega}{k^2}\left(\frac{1}{4}Tr[\slashed{K}Re(\Sigma')]\right)\,,\\
    &C=-\frac{1}{4m_{e}}Tr[Re(\Sigma')]\,,
\end{split}
\label{eq:abc}
\end{equation}
with $\Sigma'$ the $T$-dependent part of the fermion self-energy and $k=|\bk|$. In particular, the traces in Eq.~\eqref{eq:abc} are computed as integrals over the quadri-momentum of the particle in the loop. After simple integrations over the energy $p^{0}$ and the angular variables, one obtains an integral over the magnitude of the momentum $p=|\bp|$
\vspace{1cm}
\begin{widetext}
\begin{equation}
\begin{split}
   & \frac{1}{4}Tr[Re(\Sigma')]=\frac{\alpha m_{e}}{\pi k}\int_0^{\infty} dp \left[ \frac{p}{\sqrt{p^2+m_{e}^2}}(L_{2,+}n_{F,-}+L_{2,-}n_{F,+})-n_B(L_{1,+}+L_{1,-})\right]\\
    &\frac{1}{4}Tr[\slashed{K}Re(\Sigma')]=\frac{\alpha}{2\pi}\int_0^{\infty} dp \left(4p+\frac{\omega^2-k^2+m_{e}^2}{2k}(L_{1,+}+L_{1,-})\right)n_B
    + \frac{p}{\sqrt{p^2+m_{e}^2}}\left(2p-\frac{\omega^2-k^2+m_{e}^2}{2k}L_{2,+}\right)n_{F,-}\\
    & \quad\quad\quad\quad\quad\quad +\frac{p}{\sqrt{p^2+m_{e}^2}}\left(2p-\frac{\omega^2-k^2+m_{e}^2}{2k}L_{2,-}\right)n_{F,+}\\
    &\frac{1}{4}Tr[\slashed{u}Re(\Sigma')]=\frac{\alpha}{2\pi k}\int_0^{\infty} dp \{[\omega (L_{1,+}+L_{1,-})+p (L_{1,+}-L_{1,-})] n_{B}+ p (L_{2,+}n_{F,-} - L_{2,-}n_{F,+})\}\,,
    \end{split}
\end{equation}
\end{widetext}
where $L_{1,\pm}$ and $L_{2,\pm}$ are logarithmic functions arising from the integration over the angular variable, given by
\begin{equation}
\begin{aligned}
    L_{1,\pm}&=\pm\log\frac{p\,(\omega+k)\pm(\omega^2-k^2-m_{e}^2)/2}{p\,(\omega-k)\pm(\omega^2-k^2-m_{e}^2)/2}\,,\\
    L_{2,\pm}&=\pm \log\frac{\sqrt{p^2+m_{e}^2}\omega+p\,k\pm(\omega^2-k^2+m_{e}^2)/2}{\sqrt{p^2+m_{e}^2}\omega-p\,k\pm(\omega^2-k^2+m_{e}^2)/2}\,,
\end{aligned}
\end{equation}
while $n_B$ is the Bose-Einstein distribution function and $n_{F,\pm}$ the Fermi-Dirac distribution for fermions (-) and antifermions (+).
In this way, the most general dispersion relation for an electron with bare mass $m_{e}$ at finite temperature $T$ and non-vanishing chemical potential $\mu_e$ is obtained.\\
Since $m_e\ll T$ for typical SN conditions, $A,\,B,\,C$ can be assumed real~\cite{Petitgirard:1991mf}. 

\begin{figure}[t!]
\centering
\includegraphics[width=0.45\textwidth]{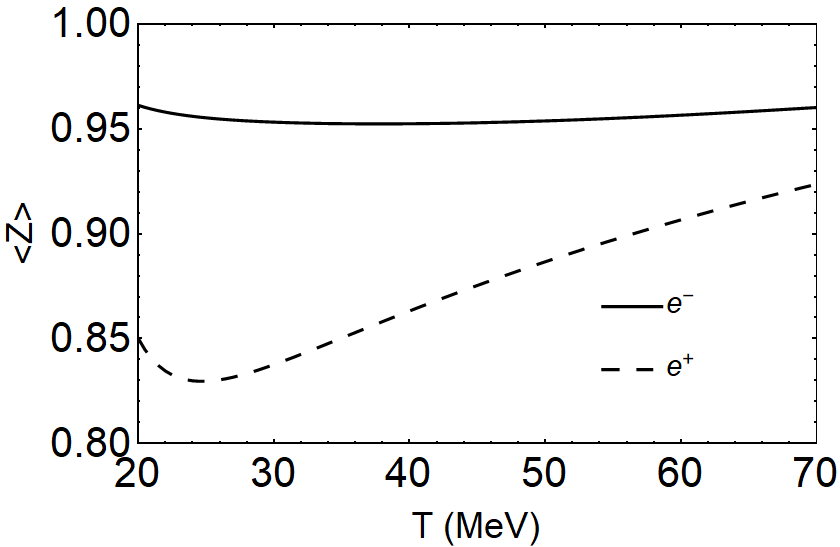}
\caption{Averaged electron and positron renormalization factors as function of the temperature for $\mu_e=200$~MeV.}
\label{fig:zemean}
\end{figure}

\begin{figure}[t!]
\centering
\includegraphics[width=0.45\textwidth]{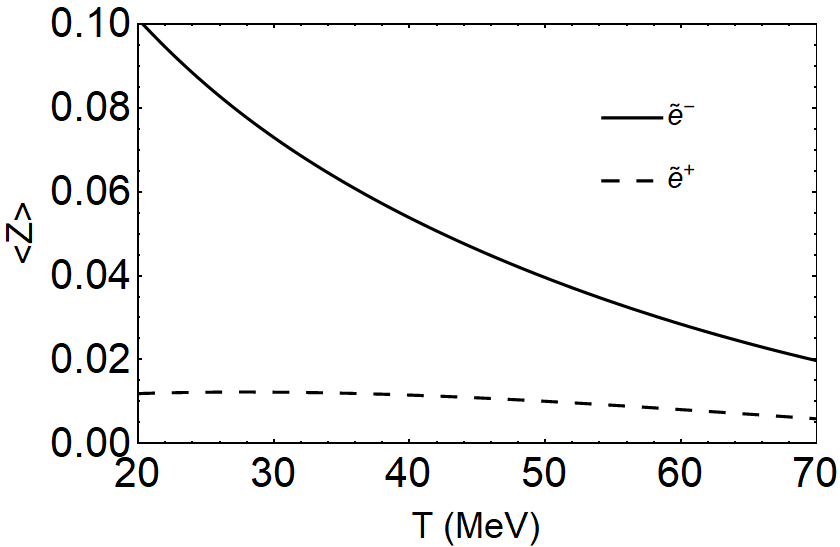}
\caption{Averaged plasmino and antiplasmino renormalization factors as function of the temperature for $\mu_e=200$~MeV.}
\label{fig:zpmean}
\end{figure}

As pointed out in Ref.~\cite{Braaten:1991hg}, Eq.~\eqref{eq:trasceq} has four solutions: two couples of solutions almost symmetric around zero for each momentum. The couple of positive energies at higher and lower energies represents the electron and plasmino excitation, respectively. The plasmino is a collective mode of the plasma that disappears in the high-momentum limit ($k/T\gtrsim 0.5$). This behavior already suggests that extra-modes introduced by thermal effects in a SN plasma are not relevant for the typical momenta close to the Fermi energy for electrons and to the temperature for positrons. The couple of solutions at negative energies are the corresponding antiparticles. The particle and antiparticle energies are not completely equal, reflecting the unequal presence of electrons and positrons. Practically, these differences are smaller (or even much smaller) than $\sim 5\%$ for the range of interest and would disappear completely for a vanishing electron chemical potential.

In the ultra-relativistic and degenerate SN plasma, the electron energy $E_{e^-}$ is of the order of the Fermi energy or chemical potential $\mu_e$, $E_{e^-}\sim \mu_e \sim O(200)\, \MeV$; while the positron energy $E_{e^+}$ is comparable with the environment temperature $T$, $E_{e^+}\sim T$.
Finding the poles of the propagator, one obtains the dispersion relations of electrons (solid black line) and positrons (dashed red line) in Fig.~\ref{fig:disprel}.  In the SN conditions, as shown in Fig.~\ref{fig:disprel}, the $e^-$ and $e^+$ dispersion relations are well approximated (for high momenta) by \cite{Braaten:1991hg}
\begin{equation}
\omega^{2}=k^{2}+2(m_{e}^{*})^{2}\,,
\label{eq:disp}
\end{equation}
where $m_e^*$ is the effective electron mass
\begin{equation}
m_e^*=m_e/2+(m_e^2/4+M^2)^{1/2}\,, 
\label{eq:meff}
\end{equation}
with $M^{2}=e^{2}(\mu_e^{2}+\pi^{2}T^{2})/8\pi^{2}$. The advantage of this result, valid in the high momentum limit, is the minimal modification of the free-electron dispersion relation in a degenerate and relativistic plasma, which requires just the replacement $m_e\rightarrow \sqrt{2} m_e^*$.\\
In this context, the coupling of the particle and hole excitations to the plasma is rescaled through a renormalization factor, which can be evaluated as~\cite{Weldon:1989ys}
\begin{equation}
\begin{split}
D(\omega,k)&=(1+A)^2(\omega^{2}-k^{2})+\\
&\quad+2(1+A)B\,\omega+B^2-m_{e}^2(1-C)^2\,,\\
    Z_{l}^{-1}&=\frac{\partial D(\omega,k)}{\partial\omega}\bigg|_{\omega=E_{l}}\,,
    \end{split}
\end{equation}  
where $l$ can represent the electron, positron or plasminos. In Fig.~\ref{fig:zemean} the thermal averaged electron and positron renormalization factors as a function of the temperature $T$ at fixed value of the chemical potential $\mu_e=200$ MeV is shown. Since $E_{e^-}\sim \mu_e$, the electron renormalization factor $Z_{e^-}\approx1$, with a weak $T$-dependence. On the other hand, since $E_{e^+}\approx T$, $Z_{e^+}<Z_{e^-}$ and it increases with $T$. However, even the positron renormalization factor deviates less than $15\%$ from unity.\\
\begin{figure}[t!]
\centering
\includegraphics[width=0.55\textwidth]{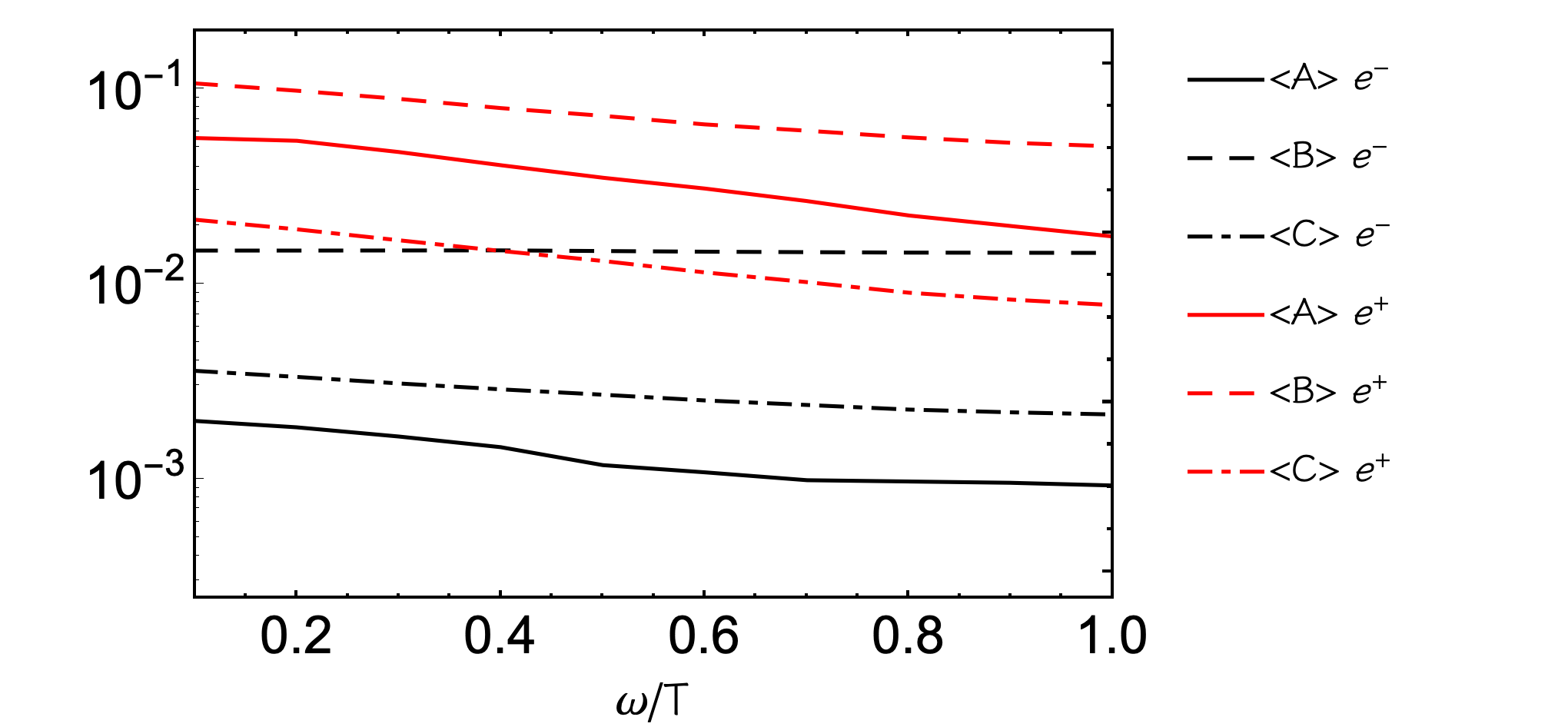}
\caption{Averaged electron and positron $A$, $B$ and $C$ functions as function of the fermion energy for $\mu_e=200$~MeV and $T=30$~MeV.}
\label{fig:abc}
\end{figure}
By contrast, as shown in Fig.~\ref{fig:zpmean}, the averaged renormalization factors for plasminos are lower than 0.1, as discussed in Ref.~\cite{Braaten:1991hg}. Therefore, the effects of the plasmino pole disappear at high momentum ($E\approx k\sim T\gg M$), consistent with the identification of the plasmino as a collective mode of the plasma like the plasmon. For this reason, we will ignore the plasmino contribution to ALP production.\\

\subsection{ALP-fermion interaction in a relativistic plasma}
The plasma effects in a SN could have impact also on the ALP-electron coupling. Indeed, the ALP interaction with electrons is described by the following Lagrangian
\begin{equation}
    \mathcal{L}_{ae}=\frac{g_{ae}}{2 m_e} \bar{\psi}_{e}\gamma^{\mu}\gamma^{5}\psi_{e} \partial_\mu a\,,
\label{eq:lagrder}
\end{equation}
where $\psi_e$ and $a$ are, respectively, the electron and ALP fields, $m_e$ is the electron mass, and $g_{ae}$ is the dimensionless ALP-electron coupling. In the case of free-electron theories, exploiting the Dirac equation $(i\slashed{\partial}-m_e)\psi_e=0$, one can prove that the Lagrangian in Eq.~\eqref{eq:lagrder} is equivalent to
\begin{equation}
  \mathcal{L}_{ae}=  -i g_{ae}\bar{\psi}_{e}\gamma^{5}\psi_{e}\,a\,.
 \label{eq:pseudo}
\end{equation}
We stress that this equivalence is not general; e.g. it ceases to be valid when two Goldstone bosons are attached to one fermion line~\cite{Turner:1988bt,Choi:1988xt}. This equivalence was exploited in Ref.~\cite{Carenza:2021osu} to revise the ALP production via electron-ion bremsstrahlung. Below, we show that the equivalence between the Lagrangians in Eq.~\eqref{eq:lagrder} and~\eqref{eq:pseudo} still holds in the degenerate and relativistic SN plasma. Applying the Dirac equation in Eq.~\eqref{eq:dirac}, the derivative coupling in Eq.~\eqref{eq:lagrder} is found to be equivalent to
\begin{equation}
    \bar{\psi}\gamma^\mu \gamma^5 \psi \partial_\mu a = - 2 \mathcal{R}\left[\frac{i m_{e} (1-C) \bar{\psi} \gamma^5 \psi a - i B \bar{\psi} \slashed{u} \gamma^5 \psi a}{1+A}\right]\,,
\label{eq:equiv}
\end{equation}
which reduces to Eq.~\eqref{eq:pseudo} in the vanishing limit of $A,\,B,\,C$. As shown in Fig.~\ref{fig:abc}, the averaged values of these quantities for typical SN conditions are lower than $10^{-1}$ for positrons (red lines) and even smaller $10^{-2}$ for electrons (black lines). Therefore, considering the uncertainties related to the SN conditions, the derivative and the pseudoscalar couplings would be considered approximately equivalent in the limit $A,\,B,\,C\ll O(1)$.
Here we summarize the results obtained in this Section:
\begin{itemize}
    \item The contribution of the (anti)plasmino in scattering processes is negligibly small due to the behavior of the renormalization factor at the typical energies.
    \item Since $E_{e^-}\sim \mu_e$ and $E_{e^+}\sim T$, the renormalization factors of electrons and positrons are close to one.
    \item For such large energies, also the equivalence between pseudoscalar and derivative ALP couplings is preserved.
    \item The fermion dispersion relation is similar to the free particle case with only a change in the mass.
\end{itemize}
Thus, to compute the ALP emissivity, it is possible to use the standard recipe shown in Sec.~\ref{sec:production} with the mass replacement $m_{e}\rightarrow \sqrt{2}\,m_{e}^{*}$.
 
\section{ALP production mechanisms}
\label{sec:production}

\subsection{Electron-ion bremsstrahlung}

This Section shortly introduces the electron-ion bremsstrahlung extensively discussed in Ref.~\cite{Carenza:2021osu}. This process consists in the interaction of an electron with a ion electric field, and the final electron emits an ALP. In this context, given the equivalence between Eqs.~\eqref{eq:lagrder} and \eqref{eq:pseudo} valid also in a SN within the limits discussed in Sec.~\ref{sec:disprel}, the electron-ion bremsstrahlung matrix element is
\begin{equation}
\begin{split}
    &\mathcal{M}_{j}=\frac{g_{ae}\,Z_{j}e^{2}}{|\bq|(|\bq|^{2}+k_{S}^{2})^{1/2}}\\
    &\times\bar{u}(p_{f})\left[\gamma^{5}\frac{1}{\slashed{P}-m_{e}}\gamma^{0}+\gamma^{0}\frac{1}{\slashed{Q}-m_{e}}\gamma^{5}\right]u(p_{i})\,,
\end{split}
\label{eq:matel}
\end{equation}
where $u(p_{i}), u(p_{f})$ are the electron spinors, $p_{i}$, $p_{f}$ and $p_{a}$ are four-momentum of initial, final electrons and ALP, $P=p_{f}+p_{a}$, $Q=p_{i}-p_{a}$, and $\bq=\bp_{f}+\bp_{a}-\bp_{i}$ is the momentum transfer. The term $[|\bq|(|\bq|^{2}+k_{S}^{2})^{1/2}]^{-1}$ is the Coulomb propagator in a plasma and $k_{S}$ is the Debye screening scale given by \cite{Raffelt:1985nk}
\begin{equation}
k_S^2 = \frac{4\pi \alpha \sum_j Z_j^2 n_j}{T}\,,
\end{equation}
where $n_j$ the number density of ions with charge $Z_j\,e$ and $\alpha$ is the fine structure constant.
The ALP flux is found to be 
\begin{equation}
\begin{split}
    \frac{d^{2}n_{a}}{dt\,d\omega_{a}}=&2\pi\int\frac{2d^{3}\bp_{i}}{(2\pi)^{3}2E_{i}}\frac{2d^{3}\bp_{f}}{(2\pi)^{3}2E_{f}}\frac{|\bp_{a}|}{(2\pi)^{3}}\\
    &(2\pi)\delta(E_{i}-E_{f}-\omega_{a})\,|\mathcal{M}|^{2}f_{i}(1-f_{f})=\\
    =&\frac{1}{64\pi^{6}}\int d\cos\theta_{ia}\,d\cos\theta_{if}\,d\delta\,dE_{f}\\
    &|\bp_{i}||\bp_{f}||\bp_{a}||\mathcal{M}|^{2} f_{i}(1-f_{f})\,,
\end{split}
\label{eq:flux}
\end{equation}
where $\omega_{a}$, $E_{i}$ and $E_{f}$ are the energies of the ALP, initial and final electrons respectively; $f_{i,f}$ are the electron distribution functions; $\theta_{ia}$, $\theta_{if}\in [0,\pi]$ are the angles between the initial electron and the ALP and the final electron moments respectively; $\delta\in[0,2\pi]$ is the angle between the two planes determined by the vectors $\bp_i-\bp_a$ and $\bp_i-\bp_f$ and $|\mathcal{M}|^{2}=\frac{1}{4}\sum_{j} n_j \sum_{s} |\mathcal{M}_j|^{2}$ is the matrix element in Eq.~\eqref{eq:matel} averaged over the electron spins and summed over all the target ions. 
The exact form of this matrix element is given in the Appendix~A of Ref.~\cite{Carenza:2021osu}. 
For sake of clarity, we write here the matrix element for a vanishing ALP mass
\begin{equation}
\begin{split}
 &|\mathcal{M}|^{2}=\frac{1}{4}\sum_{j}n_{j}\sum_{\rm s}|\mathcal{M}_{j}|^{2}=\frac{g_{ae}^{2}e^{2}}{2}\frac{k_{S}^{2}T}{|\bq|^{2}(|\bq|^{2}+k_{S}^{2})}\\
 &\left[2\omega_{a}^{2}\frac{p_{i}\cdot p_{f}-m_{e}^{2}-q\cdot p_{a}}{(p_{i}\cdot p_{a})(p_{f}\cdot p_{a})}+2-\frac{p_{f}\cdot p_{a}}{p_{i}\cdot p_{a}}-\frac{p_{i}\cdot p_{a}}{p_{f}\cdot p_{a}}\right]\,,
  \end{split}
  \label{eq:matel2}
\end{equation}
where $q=p_{f}-p_{i}$. \\
We remark that the calculation above is valid in a strongly interacting plasma as long as the equivalence between Eqs.~\eqref{eq:lagrder} and \eqref{eq:pseudo} holds and the appropriate electron dispersion relation, modified by thermal effects, is taken into account. Indeed, as discussed in Sec.~\ref{sec:disprel}, electrons acquire an effective mass in a SN core. For this reason, from now on we will naively apply Eq.~\eqref{eq:flux}, by replacing the bare electron mass with the effective one $m_e\rightarrow \sqrt{2} m_e^*$, following Eq.~\eqref{eq:disp}.\\
In particular, in order to evaluate the impact of the ALP production via electron bremsstrahlung in a SN, one can compute the ALP emissivity, i.e. the energy emitted per unit mass and time, as
\begin{equation}
    \varepsilon_a=\frac{1}{\rho}\int_{m_a}^\infty d\omega_a \omega_a \frac{d^2 n_a}{dtd\omega_a}\,,
\end{equation}
where $\rho$ is the matter density and $d^2 n_a/dt d\omega_a$ is evaluated from Eq.~\eqref{eq:flux}, taking $Z=1$ (since electrons interact with the free protons electric field in the SN core) and $n=\rho Y_e/m_N$, with $Y_e$ the electron fraction and $m_N=938$~MeV the nucleon mass. In the following, the main features of the electron bremsstrahlung in the SN core will be analyzed by assuming a schematic SN model with constant representative values for temperature $T=30$~MeV, density $\rho=3\times10^{14}$ g cm$^{-3}$ and electron fraction $Y_e=0.3$ \cite{Raffelt:1996wa}, to which we refer as \textit{typical SN conditions}. For these conditions, the effective electron mass and the electron Fermi energy are $m_e^*=8.7~\MeV$ and $E_F\approx 230$~MeV, respectively.

\subsection{Mass suppression of the bremsstrahlung}
\label{sec:suppression}

\begin{figure}
\centering
\includegraphics[width=0.45\textwidth]{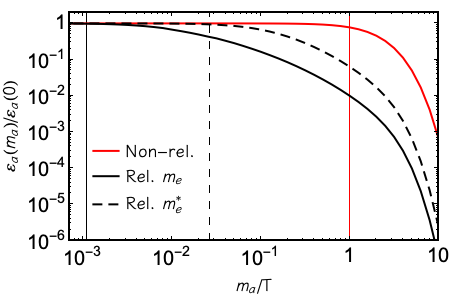}
\caption{Ratio $\varepsilon_{a}(m_{a})/\varepsilon_{a}(0)$ as function of the ALP mass for non-relativistic (red line) and relativistic conditions (black lines). The solid black line is obtained by considering the bare electron mass; the dashed one is calculated including the effective electron mass $m_{e}^{*}=8.7\,\MeV$. The vertical lines indicate the beginning of the mass suppression: $m_{a}=T$ for the non-relativistic case (red),  $m_{a}=m_{e}T/2 E_{F}$ for the relativistic cases, where $m_{e}$ can be the bare electron mass (solid black) or $\sqrt{2}\,m_e^*$, if the effective electron mass in considered (dashed black).}
\label{fig:Qratio}

\end{figure}

\begin{figure}[t!]
\centering

\includegraphics[width=0.45\textwidth]{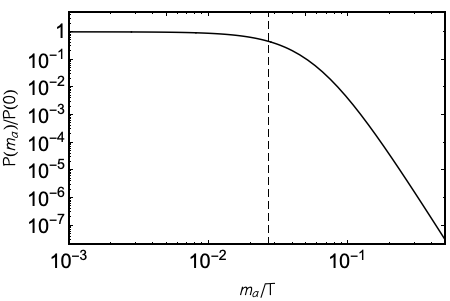}
\caption{Suppression factor $P(m_a)/P(0)$ as a function of the ALP mass for typical SN conditions, where $m_e^*\approx 8.7$~MeV. The vertical dashed line corresponds to the threshold $m_a=m_e^* T/\sqrt{2} E_F$.}
\label{fig:Ssuppr}
\end{figure}

An increase in the ALP mass determines a suppression of the emissivity. In particular, in Ref.~\cite{Carenza:2021osu} we showed that in conditions where the electron plasma is non relativistic ($T\ll m_e$), such as the Sun or red giants, the ALP emissivity is suppressed as the mass increases, due to the Boltzmann factor $e^{-m_a/T}$ in Eq.~\eqref{eq:flux}. On the other hand, the situation is strongly different in the relativistic and degenerate SN plasma ($T,\mu_e \gg m_e$). In Fig.~\ref{fig:Qratio} we show the ratio between the massive and the massless ALP emissivity as a function of the ratio between the ALP mass and temperature $m_a/T$ for  non-relativistic (red line) and relativistic (black lines) plasma conditions. In the non-relativistic case, solar conditions~\cite{Raffelt:1996wa} ($T=1.3\,\keV$, $\rho=1.6\times10^{2}$~g cm$^{-3}$, $Y_e=0.5$) have been taken as a benchmark, while typical SN conditions have been considered in the other case. In this context, it is necessary to include plasma effects on the propagation of electrons (and positrons), as described in Sec.~\ref{sec:disprel}. 
\\
As shown in Fig.~\ref{fig:Qratio}, in non-relativistic conditions, where $T\ll m_e$, the emissivity is Boltzmann suppressed at masses $m_a\gtrsim T$. On the other hand, in relativistic plasma conditions the emissivity starts to decrease for masses $m_a\ll T$, precisely when $m_{a}\sim m_{e}^{*} \omega_a/\sqrt{2}E$, where $\omega_a\sim T$ is the ALP energy and the electron energy is approximately equal to the Fermi energy $E\sim E_{F}$.  
This counter-intuitive behavior can be explained by looking at the propagator-related term  $P$ in the complete matrix element of the ALP bremsstrahlung, shown in Appendix~A in Ref.~\cite{Carenza:2021osu}:
\begin{equation}
    P=\frac{1}{(2(p_a\cdot p_f)+m_a^2))^2(m_a^2-2(p_a\cdot p_i))^2}\,,
\label{eq:prop}
\end{equation}
where $i,\,f$ stand for the initial and final electron states, respectively.
In the case of relativistic electrons and light ALPs ($m_a\ll T$), Eq.~\eqref{eq:prop} can be approximated as
\begin{equation}
    P\simeq \frac{1}{16\,\omega_a^4\,E_i^2\,E_f^2 (1-\beta_i\,\beta_a\,\cos\theta_{af})^2(1-\beta_f\beta_a\cos\theta_{ai})^2}\,,
\end{equation}
where $E_{i,f}$ is the electron energy, $\beta_{i,f}=\sqrt{1-\frac{m_e^{2}}{E^2}}$ and $\beta_{a}=\sqrt{1-\frac{m_a^2}{\omega_a^2}}$ are the electron and ALP velocities, respectively. A further simplification can be done by observing that the Coulomb scattering is mostly forward~\cite{Raffelt:1996wa} and then $\cos\theta_{ai}\simeq\cos\theta_{af}\simeq 1$, obtaining
\begin{equation}
 P\simeq\frac{1}{16\omega_{a}^{4}E_{i}^{2}E_{f}^{2}(1-\beta_{i}\beta_{a})^{2}(1-\beta_{f}\beta_{a})^{2}}\,.
     \label{eq:rel}
\end{equation} 
Since in a SN the electrons are relativistic and degenerate, assuming $E_i\simeq E_f\simeq E_F$, where $E_F$ is the electron Fermi energy, and $\omega_a\simeq T$, Eq.~\eqref{eq:rel}, expanded to the second order in $m_a/T$, becomes
\begin{equation}
    P\simeq \frac{1}{16\,E_F^4\,T^4\,(1-\beta_F)^4}-\frac{m_a^2 \beta_F}{8E_F^4\,(1-\beta_F)^5\,T^6}\,,
    \label{eq:relappr}
\end{equation}
where $\beta_F$ is the Fermi velocity. As shown in Fig.~\ref{fig:Ssuppr} for typical SN conditions, from Eq.~\eqref{eq:relappr} the $P$ term starts to decrease when
\begin{equation}
    m_a^2\gtrsim (1-\beta_F)\frac{T^2}{2\beta_F}\,,
\end{equation}
which expanded at the second order in $m_{e}/E_F$ gives
\begin{equation}
    m_a\gtrsim \frac{m_e\,T}{2\,E_F}\rightarrow\frac{m_e^{*}\,T}{\sqrt{2}\,E_F}\,,
\end{equation}
where in the last step we have replaced $m_{e}\rightarrow \sqrt{2}m_{e}^{*}$.
In this sense, the effective electron mass is an additional energy scale which becomes relevant in a SN, where electrons are relativistic and degenerate. This behavior is confirmed by Fig.~\ref{fig:Qratio}, where we compare the suppression in relativistic conditions for two different electron masses, namely the bare electron mass (black solid line) and the effective electron mass in a plasma (black dashed line). As discussed, the ALP emissivity is suppressed for $m_{a}\gtrsim m_{e}^{*}T/\sqrt{2}E_{F}$, depending linearly on $m_{e}^{*}$. Thus, the suppression begins at higher ALP masses as the electron mass increases. We stress that the bare electron mass is shown in Fig.~\ref{fig:Qratio} just for pedagogical purposes, since in a SN the effective electron mass must be taken into account.
We remark that a similar phenomenon is reported in \cite{Ebr:2013iea} in the case of the dark-photon bremsstrahlung, but not fully appreciated. Finally, this anomalous suppression is absent in non-relativistic conditions since in this case, when the Boltzmann suppression starts, $m_e\gg T\simeq m_a$ and the propagator-related term in Eq.~\eqref{eq:prop} can be approximated as
\begin{equation}
    P\simeq \frac{1}{16\omega_a^4 m_e^4}\,,
\end{equation}
independently on the ALP mass.

\subsection{The electron-positron fusion}

\begin{figure}[t]
\centering
\includegraphics[width=0.45\textwidth]{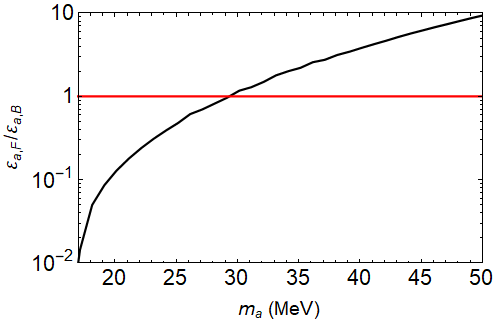}
\caption{Ratio of the electron-positron fusion and bremsstrahlung emissivities as function of the ALP mass for typical SN conditions.}
\label{fig:ratio}
\end{figure}

For ALP masses $m_a\lesssim 30$~MeV, the main production channel in a SN is the bremsstrahlung on free protons $p$,  $e^{-}+p\rightarrow e^{-}+p+a$ \cite{Raffelt:1996wa}, 
while the Compton scattering $e^{-}+\gamma\rightarrow e^{-}+a$ and the electron-positron photonic annihilation $e^{+}+e^{-}\rightarrow \gamma+a$ are suppressed by electron degeneracy \cite{Pantziris:1986dc,Raffelt:1996wa}, and even more by the ALP mass.
At ALP masses $m_a > 2 m_e$ a new process might compete with the electron-ion bremsstrahlung: the electron-positron fusion $e^{+}+e^{-}\rightarrow a$. The matrix element of this process is given by
\begin{equation}
 \frac{1}{4}\times\sum|\mathcal{M}|^{2}=\frac{g_{ae}^{2}m_{a}^{2}}{2}\,;
\end{equation}
and the ALP spectrum is obtained by integrating over the electron and positron distributions
\begin{equation}
    \frac{d^{2}n_{a}}{dt\,d\omega_{a}}=\frac{g_{ae}^{2}m_{a}^{2}}{16\pi^{3}}\int_{E_{min}}^{E_{max}} dE_{+}\, f_{+}f_{-}\,;
\end{equation}
where $E_{\mp}$ and $\bp_{\mp}$ are the electron/positron energy and momentum, $f_{\mp}$ are the electron/positron distributions and due to the energy-momentum conservation $2\omega_{a}E_{+}=m_{a}^{2}+2\bp_{a}\cdot \bp_{+}$, implying the limits of integration
\begin{equation}
E_{min,max}=\frac{\omega_{a}}{2}\mp\frac{\sqrt{\omega_a^2\,m_a^2-m_a^4-4|\bp_a|^2\,m_e^2}}{2m_a} \,.   
\end{equation}
As shown in Fig.~\ref{fig:ratio} for typical SN conditions, the electron-positron fusion becomes the dominant process at ALP masses $m_{a}\gtrsim 30\,\MeV$. This threshold strongly depends on the effective electron mass $m_e^*$, evaluated through Eq.~\eqref{eq:meff}, therefore it is strictly related to the SN properties determining $m_e^*$.

\section{The SN 1987A bound}
\label{sec:SNbound}

\begin{figure}[t]
\centering
\includegraphics[width=0.45\textwidth]{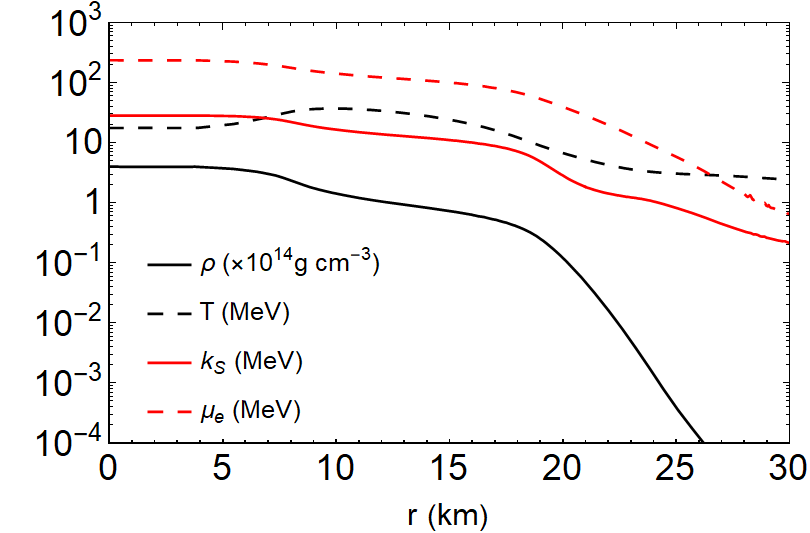}
\caption{Radial profiles of density (black solid line), temperature (black dashed line), screening scale (red solid line) and electron chemical potential (red dashed line) at $t_{pb}=1$~s for the SN model with $18~M_\odot$ progenitor mass.}
\label{fig:pl_T_profile}
\end{figure}

In the following, in order to place a bound on the ALP-electron coupling $g_{ae}$, we consider the SN simulations of Ref.~\cite{Fischer:2018kdt}. These are based on the the general relativistic neutrino radiation hydrodynamics model {\tt AGILE-BOLTZTRAN}, featuring three-flavor Boltzmann neutrino transport~\cite{Mezzacappa:1993gn,Liebendoerfer:2002xn} including a complete set of weak interactions~(see Table~I in Ref.~\cite{Fischer:2018kdt}). The equation of state used in Ref.~\cite{Fischer:2018kdt} is the modified nuclear statistical equilibrium (NSE) model of Ref.~\cite{Hempel:2009mc} for the description of heavy and light nuclei (for a recent comparison of the NSE model with the generalized density-functional approach, see Ref.~\cite{Fischer:2020}), combined with the DD2 density-dependent relativistic mean field model of Ref.~\cite{Typel:2009sy}. The SN simulations were launched from the 18~M$_{\odot}$ progenitor of the stellar evolution calculations of Ref.~\cite{Woosley:2002zz}. In spherical symmetry, neutrino-driven explosions cannot be obtained~\cite{Janka:2006fh}, except for the class of electron-capture SNe associated with progenitors of zero-age main sequence masses of about  8--9~M$_\odot$~\cite{Kitaura:2005bt,Fischer:2009af,Hudepohl:2010}. Hence, in order to trigger the neutrino-driven explosion in spherical symmetry, in Ref.~\cite{Fischer:2018kdt} the charged-current rates have been enhanced in the gain region, following the recipe described in Ref.~\cite{Fischer:2009af}. It results in the explosion onset at around 500~ms after core bounce with the continuous expansion of the SN shock to increasingly larger radii, after which the standard charged-current rates are used again.
 
\begin{figure}[t]
\centering
\includegraphics[width=0.45\textwidth]{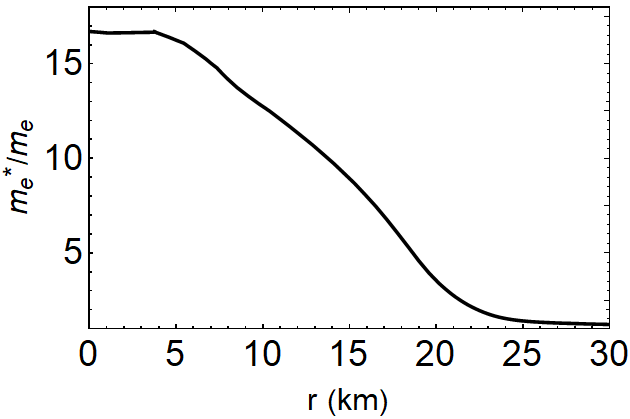}
\caption{Radial profile of the ratio between the effective and bare electron mass $m_{e}^{*}/m_e$ at $t_{pb}=1$~s for the SN model with $18~M_\odot$ progenitor mass.}
\label{fig:effmass}
\end{figure}

Here, the screening scale is computed following Ref.~\cite{Payez:2014xsa}
\begin{equation}
k_S^2 = \frac{4\pi \alpha n_p^{eff}}{T}\,;
\end{equation}
where $n_p^{eff} = 2 \int \frac{d^3p}{(2\pi)^3}f_p(1-f_p)$ is the effective number of protons in a SN core. In Fig.~\ref{fig:pl_T_profile} we show the radial profiles of some relevant quantities for the $18\,M_{\odot}$ SN model used in this paper. The effective electron mass $m_e^*$ is determined by Eq.~\eqref{eq:meff} and it is plotted in Fig.~\ref{fig:effmass}. The effective electron mass is about $\sim 15$  times larger than the bare mass in the very inner SN core ($r\lesssim 5$~km) and it decreases at larger radii.

\begin{figure}[t!]
\centering
\includegraphics[width=0.45\textwidth]{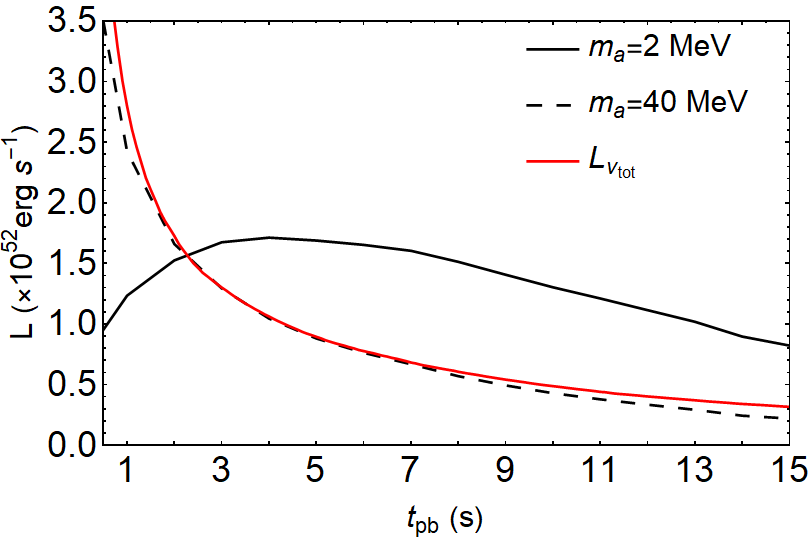}
\caption{Time evolution of the ALP luminosity for two different ALP masses $m_{a}=2\,\MeV$ (black solid line) and $m_{a}=40\,\MeV$ (black dashed line) and a coupling $g_{ae}=5\times10^{-10}$. For comparison, the total neutrino luminosity (red line) is shown.}
\label{fig:Labremm}
\end{figure}

In a SN, an additional cooling channel would ``steal'' energy to the neutrino burst, shortening its duration. As recently proposed in Ref.~\cite{Chang:2016ntp,Lucente:2020whw}, the ALP luminosity is calculated as
\begin{equation}
    L_{a}=4\pi\int_{0}^{R_\nu}dr\,r^{2}\int_{m_{a}}^{\infty}d\omega_{a}\,\omega_{a}\frac{d^{2}n_{a}}{dt\,d\omega_{a}}e^{-\tau_{a}(\omega_{a},r,R)}\,;
    \label{eq:luminosity}
\end{equation}
where $R_\nu$ is the neutrino-sphere radius, $\tau_{a}(\omega_{a},r,R)=\int_{r}^{R}\frac{d\tilde{r}}{\lambda_{a}}$ is the ALP optical depth and $\lambda_{a}$ is the ALP mean-free-path including all the possible ALP absorption processes. Only ALPs reaching a distance larger than $R$ subtract energy from the star, with $R>R_\nu$ being the so-called gain radius~\cite{Lucente:2020whw}. Depending on the value of the optical depth, we identify two different regimes for the ALP emission. If $\tau_{a}\ll1$, ALPs are in the free-streaming regime and are emitted from the entire volume of the star; if $\tau_{a}\gg1$, ALPs are in the trapping regime and are emitted as a blackbody spectrum from a spherical surface called axion-sphere.\\
In Fig.~\ref{fig:Labremm} we show the time evolution of the ALP luminosity in the free-streaming regime, for $g_{ae}=5\times10^{-10}$ and two mass values $m_a=2$~MeV (black solid line) and $m_a=40$~MeV (black dashed line), at post-bounce times $t_{pb}\gtrsim 0.5$~s. The time evolution depends on the dominant production process: the electron-ion bremsstrahlung for $m_a=2$~MeV and the electron-positron fusion for $m_a=40$~MeV. In particular, the bremsstrahlung luminosity, which is proportional to $\rho T^4$, increases in the first 3~s and then slowly decreases. On the other hand, the fusion luminosity depends only on the temperature $T$ and steeply decreases with time. As shown in Fig.~\ref{fig:Labremm}, the ALP luminosity is comparable to the total neutrino luminosity $L_{\nu_{\rm tot}}$ (red line), obtained by summing the contributions of all the (anti)neutrino flavors. It means that the ALP feedback effect on the neutrino signal cannot be ignored and should be self-consistently included in SN simulations. We postpone this task to a future work.\\
In this study, in order to constrain the ALP-electron coupling, we follow a strategy similar to the one proposed in Ref.~\cite{Raffelt:2006cw}, based on the observation of the SN 1987A neutrino burst, which requires $L_a\lesssim L_{\nu_{\rm tot}}$ for typical SN conditions. 
Specifically, we impose~\cite{Chang:2016ntp}
\begin{equation}
    L_{a}< L_{\nu_{\rm tot}}\simeq 3\times10^{52} \,\erg\,\s^{-1}\,;
    \label{eq:bound}
\end{equation}
at $t_{\rm pb}=1\,\s$. At this time, we take as benchmark values $R_\nu=21$~km and $R=24$~km~\cite{Lucente:2020whw}.

As the ALP mass or the coupling with electrons increases, it becomes more difficult for ALPs to escape freely and drain energy from the SN core. 
The only absorption processes are inverse bremsstrahlung ($a+e^{-}+p\rightarrow e^{-}+p$) and ALP decay into an electron-positron pair ($a\rightarrow e^{+}+e^{-}$). The former contributes to trapping light ALPs ($m_{a}\lesssim 30\,\MeV$) and the latter is more important for heavy ALPs. 
Indeed, both the processes produce particles that quickly thermalize in the stellar plasma and redistribute the ALP energy in the star. This means that if the ALP is absorbed in the SN core, it does not contribute to the energy-loss. With standard methods~\cite{Raffelt:1996wa} we obtained  the following ALP mean free path against inverse bremsstrahlung, important for $m_{a}\lesssim 30\,\MeV$ 
\begin{equation}
\begin{split}
    \lambda_{a,B}^{-1}=&\frac{e^{\omega_a/T}}{32\pi^{3}\omega_{a}}\int d\cos\theta_{ai}\,d\cos\theta_{af}\,d\delta\, dE_{f}\\
    &|\bp_{i}||\bp_{f}||\mathcal{M}|^{2}\,f_i(1-f_f)\,;
    \end{split}
\end{equation}
and decay, relevant for $m_{a}\gtrsim 30\,\MeV$
\begin{equation}
    \lambda_{a,D}^{-1}=\frac{g_{ae}^2}{8\pi}\,m_a\sqrt{1-\frac{4\,m_e^2}{m_a^2}}(\beta\,\gamma)^{-1}\,;
\end{equation}
where $\gamma=\omega_a/m_a$ and $\beta=\sqrt{1-\gamma^{-2}}$. Thus the total mean free path is $\lambda_{a}^{-1}=\lambda_{a,B}^{-1}+\lambda_{a,D}^{-1}$. 
\begin{figure}[t!]
\centering
\includegraphics[width=0.45\textwidth]{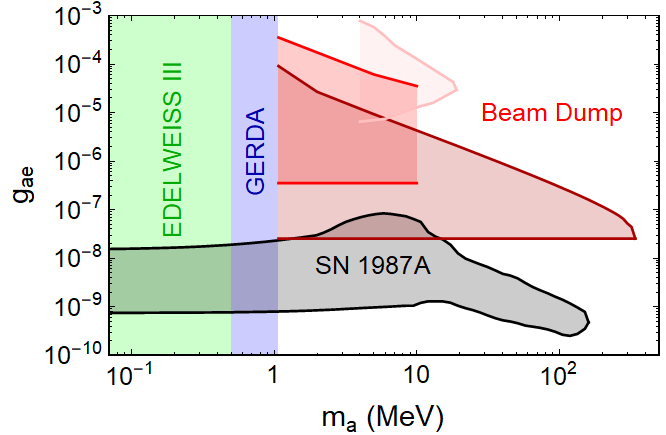}
\caption{The SN 1987A bound computed in this paper (grey) compared with beam-dump bounds (reddish), the EDELWEISS III bound (green) and the GERDA bound (blue). [The beam-dump bounds are taken from \cite{Riordan:1987aw,Bross:1989mp} (pink), \cite{Konaka:1986cb} (red) and \cite{Bjorken:1988as} (dark red).]}
\label{fig:bound}
\end{figure}
By requiring that the luminosity in Eq.~\eqref{eq:luminosity} satisfies the constraint in Eq.~\eqref{eq:bound}, the SN 1987A bound can be computed. 
The resulting bound is shown in Fig.~\ref{fig:bound} (grey) and compared with experimental bounds. A word of caution is needed here. The bound obtained in this work from Eq.~\eqref{eq:bound} is a theoretical bound, which is not on the same footing as a limit obtained from experimental data. Indeed, experimental bounds are the result of a solid statistical analysis and statements about the statistical strength of the limit can be made through the confidence level (C.L.). This is not the case for our theoretical bound, affected for instance by uncertainties on the SN structure and dynamics. In Fig.~\ref{fig:bound}, the reddish regions are excluded through beam-dump experiments \cite{Riordan:1987aw,Bross:1989mp,Bjorken:1988as,Konaka:1986cb,Alves:2017avw,Bassompierre:1995kz,Scherdin:1991xy,Tsai:1989vw}, valid at $m_a>2m_e$, in which ALPs are produced by the bremsstrahlung of an electron beam on target nuclei and then decay in flight into detectable electron-positron pairs. The vertical lines in the pink \cite{Riordan:1987aw,Bross:1989mp} and red \cite{Konaka:1986cb} regions (excluded at $90\%$ C.L.) are due to the lack of data in the original papers. The strongest beam-dump bound (dark red), given in Ref.~\cite{Bjorken:1988as}, is a $95\%$ C.L. limit in the range 1-300 MeV and it is rescaled to account for very short lived ALPs that cannot be observed. The other bounds shown in Fig.~\ref{fig:bound} are $90\%$ C.L. limits from EDELWEISS III \cite{Armengaud:2018cuy} (green) and GERDA \cite{GERDA:2020emj} (blue), which assume that ALPs constitute the whole dark matter. In these experiments, the electron recoil generated in the inverse electron bremsstrahlung might give a detectable signal in an array of cryogenic germanium detectors. The non-observation of any signal in EDELWEISS III and GERDA allows one to exclude values of the ALP-electron coupling  $g_{ae}\gtrsim 4\times10^{-11}$ close to $m_{a}\sim m_{e}$ and $g_{ae}\gtrsim 10^{-11}$ close to $m_{a}\sim 2 m_{e}$, respectively. Analogous experiments \cite{Abe:2018owy,Fu:2017lfc,Akerib:2017uem,Abgrall:2016tnn,Aralis:2019nfa,Liu:2016osd} have a similar sensitivity to EDELWEISS and GERDA in the range of interest. Another bound valid at even lower ALP masses is given by Xenon100 \cite{Aprile:2014eoa}, which excludes $g_{ae}\gtrsim 2\times10^{-11}$ for $m_{a}\lesssim1\,\keV$.\\
For $m_{a}\lesssim m_{e}$, the SN 1987A bound excludes the region $7.5\times10^{-10}\lesssim g_{ae}\lesssim 1.5\times 10^{-8}$, already excluded by the EDELWEISS, GERDA and the RG bound \cite{Capozzi:2020cbu,Straniero:2020iyi}. However, for $m_{e}\lesssim m_{a}\lesssim 200\, \MeV$ the bound calculated in this paper is the strongest one, reaching $g_{ae}\sim 2.5\times10^{-10}$ at $m_a\sim 120$~MeV. 
The obtained bound is affected by uncertainties related to the choice of the SN model. Indeed, as discussed in Refs.~\cite{Lucente:2020whw,Calore:2021klc}, the SN temperature increases as the progenitor mass becomes larger. Comparing our reference model with two other different progenitors, namely an $M=11.2~M_\odot$ and a $M=25~M_\odot$ model, we estimated the uncertainty on the bound to be smaller than a factor two in the free-streaming region (lower bound) and even smaller in the trapping regime (upper bound), with weaker constraints for lighter models.

\section{Conclusions}
\label{sec:conclusions}

In this work, we have investigated the production of ALPs coupled with electrons in a SN core, extending the study of the electron-ion bremsstrahlung of Ref.~\cite{Carenza:2021osu} to the case of relativistic electron conditions. This extension has been possible after a preliminary study on the strongly degenerate and relativistic plasma in the SN core. Indeed, in a plasma the structure of the Dirac equation is modified through the introduction of three parameters $A,\,B,\,C$, which can be assumed real since the bare electron mass is much lower than the typical SN-core temperature. In this context, the modification of the Dirac equation affects the dispersion relation of electrons and positrons, which acquire an effective mass $m_e^*\sim O(10)$~MeV in the condition of interest. In addition, we have shown that in the limit in which $A,\,B,\,C\ll O(1)$ (condition fulfilled in the SN core, given the large electron and positron energies) the ALP-fermion coupling can be equally described by a derivative or by a pseudoscalar Lagrangian. 

These considerations allowed us to consistently extend the study of the electron-ion bremsstrahlung to the degenerate and relativistic electron plasma in the SN core. In this environment, we proved that in the evaluation of the bremsstrahlung rate the ALP mass cannot be neglected when it is comparable to the effective electron mass, since the emissivity starts to be suppressed for $m_{e}^*T/E_{F}\lesssim m_{a}\ll T$, in contrast with the usual Boltzmann suppression occurring for $m_a\gtrsim T$ in a non-relativistic plasma.
Due to this suppression, in a SN core the electron-ion bremsstrahlung is the dominant ALP production process for $m_a\lesssim 30$~MeV, while at larger masses a previously neglected process is found to be dominant: the electron-positron fusion.\\
This detailed analysis allowed us to evaluate for the first time the SN 1987A bound on ALPs for masses in the range $1-200$~MeV, accounting for the plasma effects and the ALP mass. This constraint is complementary to the direct detection experiments in this region. In particular, in the small mass limit $m_a<1$~MeV, the bound excludes values $10^{-9}\lesssim g_{ae} \lesssim 10^{-8}$, already constrained by EDELWEISS and GERDA. On the other hand, at larger masses $m_a\gtrsim 30$~MeV, our new bound probes regions untouched by laboratory experiments, excluding values $g_{ae}\gtrsim 2.5\times10^{-10}$ at $m_a\sim 120$~MeV.

\section*{Acknowledgements}
We warmly thank Georg Raffelt for bringing up the question on the impact of thermal effects on the axion production in a supernova plasma and stimulating our work. We acknowledge Tobias Fischer for providing data on the supernova simulations obtained with the AGILE-BOLTZTRAN code.
We thank Maurizio Giannotti, Joerg Jaeckel, M.C. David Marsh and  Alessandro Mirizzi for helpful comments and discussions.
The work of P.C. and G.L. is partially supported by the Italian Istituto Nazionale di Fisica Nucleare (INFN) through
the “Theoretical Astroparticle Physics” project and by
the research grant number 2017W4HA7S “NAT-NET:
Neutrino and Astroparticle Theory Network” under the
program PRIN 2017 funded by the Italian Ministero
dell’Università e della Ricerca (MUR).

\bibliographystyle{bibi.bst}
\bibliography{biblio.bib}

\end{document}